\pdfoutput=1
\documentclass[a4paper,11pt]{article}
\usepackage[utf8]{inputenc}

\usepackage[table]{xcolor}
\usepackage{jcappub}
\usepackage{latexsym}
\usepackage{amsmath}
\usepackage{amssymb}
\usepackage{bbm}

\usepackage{hyperref}
\hypersetup{
    colorlinks,
    linkcolor={violet},
    citecolor={teal},
    urlcolor={teal}
}
\usepackage{orcidlink}

\usepackage{ulem}
\usepackage{pdfsync}
\usepackage{epsfig}
\usepackage{epstopdf}

\usepackage{subfigure}
\usepackage{color}
\usepackage{comment}
\usepackage{slashed}
\usepackage{physics}

\def\beq{\begin{equation}}
\def\eeq{\end{equation}}
\def\baq{\begin{eqnarray}}
\def\eaq{\end{eqnarray}}

\newcommand{\be}{\begin{equation}} 
\newcommand{\ee}{\end{equation}}
\newcommand{\bea}{\begin{eqnarray}} 
\newcommand{\eea}{\end{eqnarray}}

\title{\huge Keeping the relation between the Starobinsky model and no-scale supergravity ACTive}

\keywords{Extended Starobinsky model, No-scale supergravity, ACT} 

\author[a]{Ioannis D.~Gialamas\orcidlink{0000-0002-2957-5276},}
\author[b]{Theodoros Katsoulas\orcidlink{0000-0003-4103-7937}}
\author[b]{and Kyriakos Tamvakis\orcidlink{0009-0007-7953-9816}}

\emailAdd{ioannis.gialamas@kbfi.ee}
\emailAdd{th.katsoulas@uoi.gr} 
\emailAdd{tamvakis@uoi.gr} 

\affiliation[a]{Laboratory of High Energy and Computational Physics, 
National Institute of Chemical Physics and Biophysics, R{\"a}vala pst.~10, Tallinn, 10143, Estonia}
\affiliation[b]{Physics Department, University of Ioannina, 45110, Ioannina, Greece}

\abstract{ We introduce a modification of the Starobinsky model in the form of an additional cubic Ricci scalar curvature term $\sim \alpha R^3$, scaled by a dimensionless parameter $\alpha$, with the resulting inflaton potential being the standard Starobinsky potential modified to first parametric order by an additive term. The resulting potential is identical to the potential obtained by a modification of the superpotential employed in the construction of the Starobinsky model in the framework of no-scale supergravity, thus, extending the correspondence between a class of no-scale supergravity models and modifications of the Starobinsky model. We analyze the inflationary predictions of the model and find that for $-4.2 \times 10^{-5} \lesssim \alpha \lesssim -1.9 \times 10^{-5}$, the modified Starobinsky model is in full agreement with the recent observational data from the Atacama Cosmology Telescope, for a range of $e$-folds, $N_\star = 50-60$.}

\begin{document}

\maketitle

%

\section{Introduction}
Cosmological inflation~\cite{Kazanas:1980tx, Sato:1981qmu, Guth:1980zm, Linde:1981mu}, originally proposed as an explanation for the flatness, homogeneity, and large-scale structure of the Universe, has been established as a directly testable theory through the CMB observations, with the principal observables being the spectral index $n_s$ and the tensor-to-scalar ratio $r$. Recent data from the Atacama Cosmology Telescope (ACT)~\cite{ACT:2025fju,ACT:2025tim} in combination with the first-year DESI measurements of baryon acoustic oscillations (BAO)~\cite{DESI:2024mwx} as well as CMB observations from BICEP/Keck~\cite{BICEP:2021xfz} and Planck~\cite{Planck:2018jri} turned out to give a value of the spectral index $n_s=0.9743\pm0.0034$, which is somewhat larger than the previous Planck value of $n_s=0.9651\pm 0.0044$. In this context, the recent observations have prompted a reevaluation of several inflationary models to ensure they align with the new data~\cite{Kallosh:2025rni, Aoki:2025wld, Berera:2025vsu, Brahma:2025dio, Dioguardi:2025vci,Gialamas:2025kef,Salvio:2025izr,Antoniadis:2025pfa,Dioguardi:2025mpp,Kuralkar:2025zxr,Gao:2025onc,He:2025bli,Drees:2025ngb,Haque:2025uri,Liu:2025qca}

The Starobinsky model~\cite{Starobinsky:1980te}, featuring a modification of general relativity through the presence of a quadratic Ricci scalar curvature term, with the inflaton arising as a component of gravity itself, has over the past years proven to be quite robust, being very consistent with Planck's $n_s$ and $r$ data~\cite{Planck:2018jri}. Independently of its consistency with observations, an important feature of the Starobinsky model is that it can be constructed in the framework of no-scale supergravity~\cite{Cecotti:1987sa,Ellis:2013xoa,Ellis:2013nxa}, a rather appealing framework which can accommodate vanishing vacuum energy and supersymmetry breaking~\cite{Cremmer:1983bf,Ellis:1983sf,Ellis:1984bm} (see also the review articles~\cite{Lahanas:1986uc,Ellis:2020lnc} and~\cite{Ellis:2015kqa} for phenomenological aspects of no-scale models). Nevertheless, the original Starobinsky model turns out to be disfavored by the recent ACT data~\cite{ACT:2025fju,ACT:2025tim}. Although the issue of a generalization of the Starobinsky model within various schemes has received some attention in the past~\cite{Ozkan:2014cua,Broy:2014xwa}, an immediate question arises now, namely whether the original Starobinsky model could be appropriately modified in order to be consistent with the new data. Given that, an additional question of interest arises, namely, whether the modified Starobinsky model can still be constructed within the framework of no-scale supergravity. The present article gives a positive answer to these questions. We introduce a modification of the original Starobinsky model by adding to the action an additional cubic term of the Ricci scalar curvature scaled by a new dimensionless parameter (see, e.g.,~\cite{Berkin:1990nu,Huang:2013hsb,Asaka:2015vza,Bhattacharya:2017uwi,Rodrigues-da-Silva:2021jab,Ivanov:2021chn,Rukpakawong:2025twl,Kim:2025dyi} for related works on inflation). We employ an auxiliary scalar representation, which in the Einstein frame corresponds to the canonical inflaton self-interaction through a Starobinsky-like potential, which, to first parametric order, is the original Starobinsky potential modified by an additive term. 

On the other hand, considering a modification of the superpotential employed in the framework of $SU(2,1)/SU(2)\times U(1)$ no-scale model, we arrive at a Starobinsky-like scalar potential that coincides to first parametric order with the one obtained by direct modification of the action by introducing the cubic term. Thus, an important connection is established between the extended Starobinsky model and the class of models within the framework of no-scale supergravity, a connection well known~\cite{Ellis:2013xoa,Ellis:2013nxa,Lahanas:2015jwa} for the original Starobinsky model and its no-scale supergravity constructions (see~\cite{Kallosh:2013lkr,Farakos:2013cqa,Ferrara:2013rsa,Ferrara:2013kca,Buchmuller:2013uta,Ketov:2013dfa,Ferrara:2013pla,Li:2013moa,Kallosh:2013yoa,Alexandre:2013nqa, Ferrara:2014ima,Kallosh:2014qta,Ellis:2014rxa,Hamaguchi:2014mza,Ellis:2014gxa,Chakravarty:2014yda,Ferrara:2014fqa,Dalianis:2014nwa,Buchmuller:2014pla,Kounnas:2014gda,Ellis:2014opa,Diamandis:2014vxa,Terada:2014uia,Basilakos:2015yoa,Diamandis:2015xra,Ellis:2017xwz,Ellis:2020xmk,Nanopoulos:2020nnh,Ellis:2020krl,Garcia:2023tkk,Diamandis:2024oww} for various applications of supergravity in inflation). Next, we proceed and analyze numerically the predictions of the above constructed model in the framework of slow-roll inflation, which are found to be in agreement with the recent ACT observations~\cite{ACT:2025fju,ACT:2025tim} for a comfortable parametric range.

In section~\ref{sec:the_model} we lay out the basic description of the model starting with its $F(R)$ form and ending in terms of its canonical scalar representation. In section~\ref{sec:no_scale} we consider its no-scale supergravity derivation, following two distinct prescriptions. In section~\ref{sec:inflation} we analyze the inflationary phenomenology of the model, establishing its agreement with recent observations for a well defined range of its parameters. Finally, in section~\ref{conclusions} we summarize briefly our results.


\section{The model}
\label{sec:the_model}

We start by considering the general $F(R)$ gravity action, $\mathcal{S}=M_P^2\int{\rm d}^4x \sqrt{-g}F(R)/2$, where the Einstein-Hilbert term is generalized to a function of the Ricci scalar $R$. To facilitate the analysis and eventual transition to a scalar-tensor theory, it is convenient to rewrite the action in terms of an auxiliary scalar field $\chi$. Introducing $\chi$ allows one to linearize the higher-order curvature terms, resulting in the equivalent action
\begin{equation}
{\cal{S}}= \frac{M_P^2}{2}\int {\rm d}^4x \sqrt{-g}\left[RF'(\chi)-\chi F'(\chi)+F(\chi)\right]\,.
\end{equation}
Note that the scalar representation requires $F'(\chi)>0$ and $F''(\chi)\neq 0$. The standard Starobinsky model~\cite{Starobinsky:1980te} corresponds to choosing the functional form $F(R)=R+R^2/(6M^2)$. In this work, we extend the Starobinsky model by incorporating a cubic correction to the gravitational action. Specifically, we consider
\be 
F(R)=R+\frac{1}{6M^2}R^2+\frac{\alpha}{9M^4}R^3\,,
\ee
where $\alpha$ is a dimensionless parameter and $M$ is the mass scale associated with the model. Such higher-order curvature corrections can arise naturally as quantum corrections to gravity (see~\cite{tHooft:1974toh,Birrell:1982ix}). From this point forward, we adopt the convention $M_P=1$ for notational simplicity. The action in terms of the auxiliary field $\chi$ then becomes
\be 
{\cal{S}}=\int {\rm d}^4x\sqrt{-g}\left[\frac{\Omega^2}{2}R-\frac{\chi^2}{12M^2}-\frac{\alpha}{9M^4}\chi^3\right]\,,
\ee
where  the conformal factor is defined as $\Omega^2=F'(\chi)$.
Next, we perform a Weyl rescaling $g_{ \mu\nu}\rightarrow\Omega^{-2}g_{ \mu\nu}$ to transition to the Einstein frame, and obtain the action
\be 
{\cal{S}}=\int {\rm d}^4x\sqrt{-g}\left[\frac{R}{2}-\frac{3}{4}\frac{(\nabla\Omega^2)^2}{\Omega^4}-\frac{1}{\Omega^4}\left(\frac{\chi^2}{12M^2}+\frac{\alpha \chi^3}{9M^4}\right)\right]\,.
\ee
To express the theory in terms of a canonically normalized scalar field, we first solve the relation
$\Omega^2=F'(\chi)$ for $\chi$, and subsequently introduce the scalar field $\phi =\sqrt{3/2}\ln\Omega^2$. The action takes the form
\begin{equation}
\mathcal{S} = \int {\rm d}^4x \sqrt{-g} \left[\frac{R}{2} -\frac{ \left(\partial_\mu\phi\right)^2}{2}-V_{R^3}(\phi) \right]\,,
\end{equation}
where
\begin{equation}
\label{eq:full_Pot}
 V_{R^3}(\phi)=\frac{M^2}{144\alpha^2}e^{-2\sqrt{\frac{2}{3}}\phi}\left(1-\sqrt{1+12\alpha(e^{\sqrt{\frac{2}{3}}\phi}-1)}\right)^2
 \left(1+2\sqrt{1+12\alpha(e^{\sqrt{\frac{2}{3}}\phi}-1)}\right)\,.
\end{equation}
Expanding the potential~\eqref{eq:full_Pot} to leading order in $|\alpha|\ll 1$, we recover the Starobinsky potential with an $\mathcal{O}(\alpha)$ correction
\be
\label{eq:pot_appro}
\frac{V_{R^3}(\phi)}{M^2}\simeq\frac{3}{4}\left(1-e^{-\sqrt{\frac{2}{3}}\phi}\right)^2-\frac{3\alpha}{2}e^{\sqrt{\frac{2}{3}}\phi}\left(1-e^{-\sqrt{\frac{2}{3}}\phi}\right)^3\,.
\ee
It is important to emphasize that the $\alpha$-dependence of the scalar potential in eq.~\eqref{eq:full_Pot}, which originates from the cubic curvature correction in the gravitational action, has a significant impact on the behavior of the potential in the large-field regime. Specifically, depending on the sign of the parameter $\alpha$ the Starobinsky inflationary plateau acquires a rising tail ($\alpha>0$) or exhibits runaway behavior ($\alpha<0$). This qualitative change can significantly impact the dynamics of inflation and the resulting cosmological predictions, altering the values of the tensor-to-scalar ratio, the spectral index, and its running, bringing them into agreement with the latest constraints reported in~\cite{ACT:2025fju, ACT:2025tim}.

It is important to note that, to avoid ghost-like instabilities in the graviton sector and to maintain stability of the theory the conditions $F'(R)>0$ and $F''(R)>0$ must be satisfied~\cite{Amendola:2006kh,Amendola:2006we,Sotiriou:2008rp,Appleby:2009uf}. These conditions are automatically satisfied for $\alpha\geq 0$. However, when $\alpha<0$ they impose an upper bound on the scalar field value. Specifically, to avoid violating the condition $F''(R)>0$, the field must satisfy $\phi < \phi_{\rm UV} \equiv \sqrt{3/2} \ln[1+1/(12|\alpha|)]$ ,  which also follows from requiring that the argument of the square roots in eq.~\eqref{eq:full_Pot} remains positive. This bound introduces an effective ultraviolet (UV) cutoff on the field range of the theory. Notably, for the range of
$\alpha$ values considered in this work, the bound $\phi_{\rm UV}$ lies in the extreme UV regime, well beyond the field values relevant for inflation, ensuring that the inflationary dynamics remain unaffected by this constraint. Admittedly, the subtle features arising in the large-field regime, including the behavior near $\phi_{\rm UV}$, are not captured by the approximate potential given in~\eqref{eq:pot_appro}. Nonetheless, the simplified form of the approximate potential plays a crucial role for connecting the $R^3$ correction with no-scale supergravity models and provides a convenient framework for analyzing the inflationary observables, which will be discussed in the following sections.

\section{No-scale supergravity derivation}
\label{sec:no_scale}
    
It has been established sometime ago that the standard Starobinsky model can be derived in the framework of no-scale Supergravity models. Recently also no scale models~\cite{Antoniadis:2025pfa} have been employed to account for possible deformations in the light of the ACT data~\cite{ACT:2025fju, ACT:2025tim}. The general framework of no-scale Supergravity starts with a K{\"a}hler potential of the form
\be 
{\cal{K}}=-3 \ln{(T+\bar{T}-h(S,\bar{S}))}\,,
\ee 
where $T$ and $S$ are the two chiral superfields required to match quadratic gravity. The function $h(S,\bar{S})$ is
\be 
h(S,\bar{S})=\frac{|S|^2}{3}+\cdots\,,
\ee
where the dots signify higher powers necessary to address stability issues. The bosonic Lagrangian derived in terms of a superpotential $W(S,T)$ is
\begin{align}
{\cal{L}}= &-\frac{3|\nabla T|^2}{(T+\bar{T}-h)^2}-\frac{3|\nabla S|^2}{(T+\bar{T}-h)^2}\left((T+\bar{T}-h)h_{S\bar{S}}+|h_S|^2\right)  +\frac{3\left(h_S\nabla S\cdot \nabla\bar{T}+c.c.\right)}{(T+\bar{T}-h)^2} \nonumber
\\
& +\frac{(W_T\bar{W}+c.c.)}{(T+\bar{T}-h)^2}-\frac{|W_S|^2}{3h_{S\bar{S}}(T+\bar{T}-h)^2} -\frac{|W_T|^2}{3(T+\bar{T}-h)} \nonumber
\\
& -\frac{\left(|W_T|^2|h_S|^2+\bar{W}_{\bar{T}}W_Sh_{\bar{S}}+W_T\bar{W}_{\bar{S}}h_S\right)}{3h_{S\bar{S}}(T+\bar{T}-h)^2}\,,
\end{align}   
where $W_T$ and $W_S$ are the partial derivatives of the superpotential with respect to $T$ and $S$ and $h_S,\,h_{S\bar{S}}$ the corresponding derivatives of $h$. We have considered two cases of no-scale supergravity models that arrive at the Starobinsky model as well as its cubic correction. It should be noted that, since a geometric supergravity embedding that includes the cubic correction is not yet available, although such a construction could possibly be achieved by an extension of Cecotti's procedure~\cite{Cecotti:1987sa} (see also~\cite{Hindawi:1995qa}), these models should be referred to as Starobinsky-like. Both models are based on the $SU(2,1)/SU(2)\times U(1)$ symmetric K{\"a}hler potential with $h=|S|^2/3$. 
One choice~\cite{Ellis:2013nxa,Lahanas:2015jwa} considers a superpotential $W=\sqrt{3}MS(T-1)(1-\varepsilon_1(T-1))$ and arrives at the Starobinsky-like model in the limit $S\rightarrow 0$. The other choice ~\cite{Ellis:2013xoa,Ellis:2013nxa} assumes that the modulus $T$ is fixed to $1/2$, while the superpotential is chosen to be of the Wess-Zumino form~\cite{Wess:1974tw}, i.e. $W(S)=M(S^2/2-\lambda S^3/3\sqrt{3})$.
In the following, we examine both cases, which we refer to as Prescription I and Prescription II, respectively.

\subsection{Prescription I}

\begin{figure}[t!]
    \centering
    \includegraphics[width=0.7\textwidth]{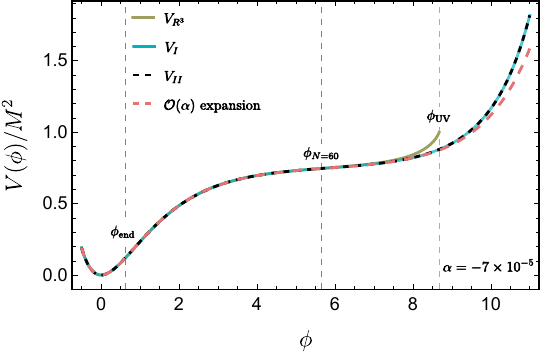}
\caption{ The scalar potentials corresponding to eqs.~\eqref{eq:full_Pot} (dark yellow), \eqref{eq:pot_I} (blue-green), and \eqref{eq:pot_II} (dashed black), along with their $\mathcal{O}(\alpha)$ expansion (dashed pink). The parameter values used are $\alpha = \varepsilon_1 = \varepsilon_2 = -7 \times 10^{-5}$.}
   \label{Fig:fig1}
\end{figure}
It is known that the superpotential choice~\cite{Ellis:2013nxa,Lahanas:2015jwa}
\be 
W(S,T)\sim S(T-1)\,,
\ee 
leads to the standard Starobinsky potential in the limit $S\rightarrow 0$, with the real part of $T$ being the inflaton. We may proceed introducing a modification~\cite{Lahanas:2015jwa}
\be 
\label{eq:superLT}
W_{I}=\sqrt{3}M S(T-1)\left(1-\varepsilon_1(T-1)\right)\,,
\ee
and investigate whether we could match the modification embodied by the cubic Ricci-curvature term. The resulting scalar potential in the $S\rightarrow 0$ limit for the real part Re$(T)=e^{\sqrt{\frac{2}{3}}\phi}$ { and Im$(T)=0$ is
\begin{equation}
 \frac{V_{I}(\phi)}{M^2}= \frac{3}{4}\left(1-e^{-\sqrt{\frac{2}{3}}\phi} \right)^2\left[1+\varepsilon_1\left(1-e^{\sqrt{\frac{2}{3}}\phi}\right) \right]^2,
\label{eq:pot_I}
\end{equation}
}
while expanding for small $\varepsilon_1$ we obtain
\be 
\frac{V_{I}(\phi)}{M^2}\simeq\frac{3}{4}\left(1-e^{-\sqrt{\frac{2}{3}}\phi}\right)^2-\frac{3\varepsilon_1}{2}e^{\sqrt{\frac{2}{3}}\phi}\left(1-e^{-\sqrt{\frac{2}{3}}\phi}\right)^3.
\label{eq:pot_I_exp}
\ee
This matches the $\frac{\alpha}{9M^4}R^3$ correction in the action for $\varepsilon_1=\alpha$.

\subsection{Prescription II}

Assuming the superpotential given in~\cite{Ellis:2013xoa,Ellis:2013nxa}, 
\begin{equation}
\label{eq:superEllis}
W_{II}(S) = M \left( \frac{1}{2} S^2 - \frac{\lambda}{3\sqrt{3}} S^3 \right)\,, 
\end{equation} and fixing the real and imaginary parts of the modulus field $T$ as Re$(T)=1/2$ and Im$(T)=0$, we can perform a field redefinition to express $S$ in terms of a canonically normalized scalar field $\phi\equiv \sqrt{6} \tanh^{-1} (S/\sqrt3)$. The scalar field potential reads
\begin{equation}
    \frac{V_{II}(\phi)}{M^2}= 3\sinh^2(\phi/\sqrt{6}) \left( \cosh(\phi/\sqrt{6}) -\lambda \sinh(\phi/\sqrt{6}) \right)^2.
\label{eq:pot_II}
\end{equation}
In the special case $\lambda = 1$, the potential~\eqref{eq:pot_II} exactly reproduces the Starobinsky inflationary potential~\cite{Ellis:2013xoa}. To study deviations from this model, we consider small perturbations around $\lambda=1$ by introducing $\lambda = 1+2\varepsilon_2$, where $\varepsilon_2\ll1$. Expanding the potential to first order in $\varepsilon_2$ we find
\begin{equation}
\label{eq:pot_Ellis_exp}
 \frac{V_{II}(\phi)}{M^2}\simeq\frac{3}{4}\left(1-e^{-\sqrt{\frac{2}{3}}\phi}\right)^2-\frac{3\varepsilon_2}{2}e^{\sqrt{\frac{2}{3}}\phi}\left(1-e^{-\sqrt{\frac{2}{3}}\phi}\right)^3\,.
\end{equation}
The potential~\eqref{eq:pot_Ellis_exp} is identical to the leading-order correction to the Starobinsky potential induced by a cubic curvature term in the effective gravitational action, given by eq.~\eqref{eq:pot_appro}, provided we identify the small parameter $\varepsilon_2$ as $\varepsilon_2= \alpha$. This model has also been discussed in the context of ACT data in~\cite{Antoniadis:2025pfa}.

The potentials given in eqs.~\eqref{eq:pot_I} and~\eqref{eq:pot_II} are completely identical for $\varepsilon_1 = \varepsilon_2$, not only at first order but at all orders. In figure~\ref{Fig:fig1}, we compare the potentials~\eqref{eq:full_Pot},~\eqref{eq:pot_I}, and~\eqref{eq:pot_II} and their first order expansion for $\alpha=\varepsilon_1=\varepsilon_2 = -7 \times 10^{-5}$. All three potentials are nearly indistinguishable in the inflationary regime ($\phi_{N=60}\simeq5.66,\, \phi_{\rm end} \simeq 0.61$). However, for larger field values, the supergravity potentials begin to deviate from~\eqref{eq:full_Pot}, which terminates at the cutoff scale $\phi_{\rm UV} \simeq 8.67$.

\begin{figure*}[t!]
    \centering
    \includegraphics[width=\textwidth]{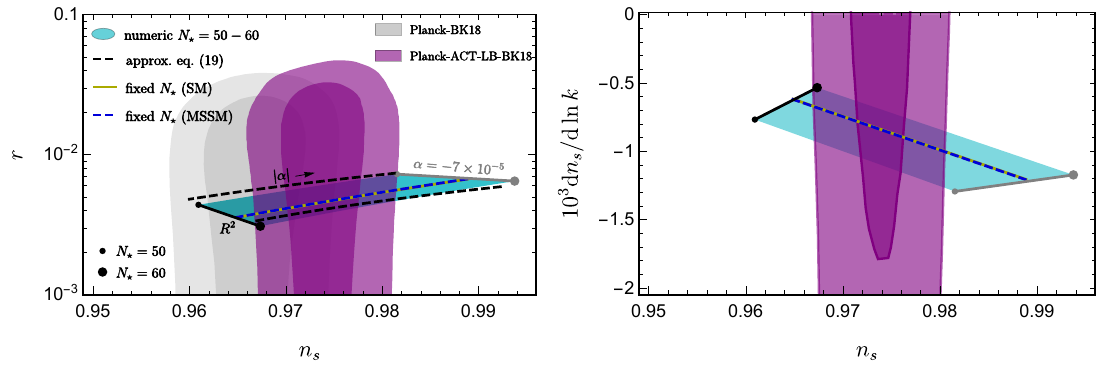}
\caption{$r$ vs $n_s$ (left) and ${\rm d}n_s / {\rm d}\ln k$ vs $n_s$ (right) at the pivot scale $k_\star = 0.05$ Mpc$^{-1}$. The gray and purple regions represent the 1$\sigma$ and 2$\sigma$ allowed areas, respectively, derived from the latest combination of Planck, BICEP/Keck, and BAO data~\cite{BICEP:2021xfz}, and from Planck, ACT, and DESI data~\cite{ACT:2025tim} respectively.
The shaded teal region illustrates the numerical results for $N_\star = 50-60$ and $\alpha=-7\times10^{-5}$ (solid black) up to $\alpha=0$ (solid gray). 
Dark yellow and dashed blue lines represent the numerical results assuming instantaneous reheating, with $N_\star$ fixed by~\eqref{eq:nstar} in the SM and MSSM, respectively. The dashed black lines show the predictions for $N_\star = 50 - 60$ based on the approximate formulas~\eqref{eq:nsr_approx} for $\alpha=-9\times10^{-5}$ up to $0$. The parameter $|\alpha|$ increases as indicated by the arrow.}
   \label{Fig:fig2}
\end{figure*}

\section{Inflation} 
\label{sec:inflation}

In this section, we analyze the inflationary dynamics of the model, constraining its parameter space by comparing its predictions to the latest observational bounds reported by ACT~\cite{ACT:2025fju, ACT:2025tim}.

We begin by introducing the Hubble slow-roll parameters, which characterize the dynamics of inflation. The first parameter is defined as $\epsilon_1 = -{\rm d}\ln H / {\rm d}N$, where $N$ is the number of $e$-folds  and $H$ the Hubble parameter. The higher-order parameters are given by $\epsilon_2 = {\rm d}\ln \epsilon_1 / {\rm d}N$ and $\epsilon_3 = {\rm d}\ln \epsilon_2 / {\rm d}N$. The amplitude of the scalar power spectrum at horizon crossing is given by $A_s^\star = H_\star^2 / (8\pi^2 \epsilon_1^\star)$, evaluated at the pivot scale $k_\star = 0.05\, {\rm Mpc}^{-1}$. Observations constrain this amplitude to $A_s^\star \simeq 2.1 \times 10^{-9}$~\cite{Planck:2018jri}, which in turn constrains the mass scale $M$ in~\eqref{eq:full_Pot}.

The scalar spectral index, which describes the tilt of the power spectrum, is given by $n_s = 1 - 2\epsilon_1 - \epsilon_2$, while its running by ${\rm d}n_s / {\rm d}\ln k = -2\epsilon_1 \epsilon_2 - \epsilon_2 \epsilon_3$. The tensor-to-scalar ratio is $r = 16\epsilon_1$. These quantities are tightly constrained by CMB observations and provide direct tests of inflationary models through data from ACT~\cite{ACT:2025fju, ACT:2025tim} and Planck~\cite{Planck:2018jri}.

Finally, the number of $e$-folds ($N_\star$) between horizon exit of the pivot scale and the end of inflation is given by~\cite{Liddle:2003as}
\be
\label{eq:nstar}
N_\star = 66.5 - \ln \left(\frac{k_\star}{a_0 H_0}\right) +\frac14\ln\left(\frac{9H_\star^4}{\rho_{\rm end}}\right)\,,
\ee
where instantaneous reheating has been assumed. The numerical coefficient $66.5$ incorporates several contributions, including the effective number of relativistic degrees of freedom during reheating, $g_{\rm reh}$. In computing this value, we have assumed a Standard Model (SM) particle content with $g_{\rm reh} = 427/4$, valid for reheating temperatures above 1 TeV. If instead the particle content corresponds to the minimal supersymmetric SM (MSSM), for which $g_{\rm reh} = 915/4$, the coefficient is slightly adjusted to $66.4$. The  Hubble constant today has been taken to be $H_0\simeq 68$ km/sec/Mpc~\cite{ACT:2025fju} and $\rho_{\rm end} = 3V(\phi_{\rm end})/2$ denotes the energy density of the inflaton field at the end of inflation.

It is worth noting that, under reasonable approximations, the scalar spectral index and the tensor-to-scalar ratio can be given by the approximate analytical expressions
\begin{equation}
\label{eq:nsr_approx}
n_s  \simeq 1-\frac{2}{N}\left[1+\frac{64\alpha} {27}N^2\right]\qquad \text{and} \qquad
r  \simeq \frac{12}{N^2} \left[1-\frac{64\alpha}{27}N^2\right]\,.
\end{equation}
It is evident that a negative sign of $\alpha$ results in an increase of the tilt $n_s$, as well as of an increase in $r$. In the limit $\alpha = 0$, one recovers the standard predictions of the Starobinsky model.

Given that the Starobinsky model tends to predict a value of $n_s$ slightly below the central value reported by ACT, thus, placing it outside the $2\sigma$ observational bounds, it is natural to explore negative values of $\alpha$. These can shift the predictions of the model into a parameter region more consistent with current data. 

\begin{figure}[t!]
    \centering
    \includegraphics[width=0.7\textwidth]{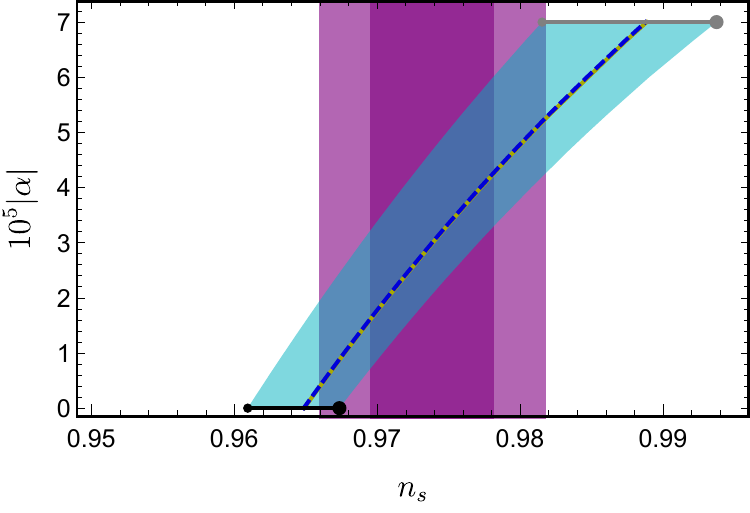}%
\caption{The parameter $\alpha\leq0$ as function of the spectral index $n_s$. The color scheme follows that of figure~\ref{Fig:fig2}. }
   \label{Fig:fig3}
\end{figure}
In our analysis, we have assumed the slow-roll approximation and we have solved numerically the Friedmann and scalar field equations using the scalar field potential~\eqref{eq:full_Pot}. The inflationary predictions are almost identical to these of the supergavity potential~\eqref{eq:pot_I} or~\eqref{eq:pot_II}.  The shaded teal region in figure~\ref{Fig:fig2} illustrates the numerical results for $N_\star = 50-60$, shown in the $r-n_s$ (left) and ${\rm d}n_s/{\rm d}\ln k-n_s$ (right) planes. The gray and purple regions represent the 1$\sigma$ and 2$\sigma$ allowed areas, respectively, derived from the latest combination of Planck, BICEP/Keck, and BAO data~\cite{BICEP:2021xfz}, and from Planck, ACT, and DESI data~\cite{ACT:2025tim}. Taking the Starobinsky model ($\alpha = 0$) as a starting point we explore negative values of $\alpha$ down to $\alpha = -7 \times 10^{-5}$. For larger values of $|\alpha|$, the spectral index becomes too large, and even $50$ $e$-folds fall outside the new bounds reported by ACT~\cite{ACT:2025fju, ACT:2025tim}. The dark yellow and dashed blue lines represent the numerical results assuming instantaneous reheating, with $N_\star$ fixed by~\eqref{eq:nstar} in the SM and the MSSM, respectively. The difference between the two models is negligible in the plot. We find that $N_\star = 55.7 - 55.9$ (SM) and $N_\star = 55.6 - 55.8$ (MSSM), while the corresponding instantaneous reheating temperatures are $T_{\rm reh} = 2.45_{-0.15}^{+0.15} \times 10^{15}$ GeV for the SM and $T_{\rm reh} = 2.05_{-0.15}^{+0.15} \times 10^{15}$ GeV for the MSSM.
The dashed black lines show the predictions for $N_\star = 50 - 60$ based on the approximate formulas~\eqref{eq:nsr_approx} for $\alpha=-9\times10^{-5}$ up to $\alpha=0$. It should be noted that, since~\eqref{eq:nsr_approx} is only an approximation, the resulting numerical predictions are reliably validated for slightly larger values of $|\alpha|$. This indicates that the approximation~\eqref{eq:nsr_approx} successfully captures the qualitative behavior of the full numerical solution. 
In the right panel of figure~\ref{Fig:fig2}, we display the running of the spectral index. Our model predicts a negative running (see also~\cite{Asaka:2015vza}), consistent with the bounds found in~\cite{ACT:2025tim}. 

Finally, in figure~\ref{Fig:fig3}, we present the value of $\alpha$ as a function of $n_s$. For $\alpha = 0$, the predictions for 60 $e$-folds are partially consistent with the data. However, for $\alpha < -7 \times 10^{-5}$, the model's predictions fall outside the allowed bounds. For $-4.2 \times 10^{-5} \lesssim \alpha \lesssim -1.9 \times 10^{-5}$, the entire region of $N_\star = 50 - 60$ is in agreement with the data. This constraint on the parameter $\alpha$ is equivalent to imposing constraints on the parameters $\varepsilon_1$ and $\varepsilon_2$ of the superpotentials~\eqref{eq:superLT} and~\eqref{eq:superEllis}, respectively.

\section{Conclusions}
\label{conclusions}

The latest data release from the Atacama Cosmology Telescope (ACT), combined with the first-year DESI measurements of BAO as well as CMB observations from BICEP/Keck and Planck, indicate a notable increase in the spectral index. This shift places the well-known Starobinsky model of inflation in significant tension with observations, at approximately the $2\sigma$ level. To resolve this, we have introduced in the action a correction term, consisting of the cube of the Ricci scalar curvature, scaled by a dimensionless parameter, namely $\sim \alpha R^3$. We find that for $-4.2 \times 10^{-5} \lesssim \alpha \lesssim -1.9 \times 10^{-5}$, the modified model becomes consistent with the updated data for a range of $e$-folds, $N_\star = 50-60$. Larger values of $|\alpha|$ increasingly constrain the available range, with $|\alpha|= 7 \times 10^{-5}$ setting the upper limit. An equally additional, important aspect of this work is the connection established between the extended Starobinsky model and a broad class of models within the framework of no-scale supergravity. It is well known~\cite{Ellis:2013xoa,Ellis:2013nxa,Lahanas:2015jwa} that the original Starobinsky model can be derived in certain no-scale supergravity constructions. We have shown that the first order {\textit{$R^3$-corrected Starobinsky model}} corresponds directly to the first-order of the appropriate no-scale supergravity constructions, with this correspondence hinting towards a deeper underlying connection.

\acknowledgments
The work of IDG was supported by the Estonian Research Council grants MOB3JD1202, RVTT3, RVTT7 and by the CoE program TK202 ``Foundations of the Universe'’. KT wishes to thank the CERN Theoretical Physics Department for hospitality.

\bibliography{references}{}

\end{document}